\documentclass{jltp}
\usepackage{graphicx}

\title{Cavitation of Electron Bubbles in Liquid Helium Below Saturation Pressure}
\author{Mart\'{\i} Pi, Manuel Barranco, Ricardo Mayol, and V\'{\i}ctor Grau }
\address{Departament ECM,
Facultat de F\'{\i}sica, Universitat de Barcelona, \\ E-08028 Barcelona,
Spain}

\runninghead{M. Pi, M. Barranco, R. Mayol, and V. Grau}
            {Cavitation of Electron Bubbles in Liquid Helium}

\begin{document}

\maketitle

\begin{abstract}

We have used a Hartree-type electron-helium potential together with a
density functional description of liquid $^4$He and $^3$He to study 
the explosion
of electron bubbles submitted to a negative pressure. The critical
pressure at which bubbles explode has been determined as a function of
temperature.  It has been found that this critical pressure is very
close to the pressure at which liquid helium becomes globally unstable
in the presence of electrons. It is shown that at high temperatures 
the capillary model overestimates the critical pressures. We have checked
that a commonly used and rather simple electron-helium interaction
yields results very similar to those obtained using
the more accurate Hartree-type interaction. We have estimated that the
crossover temperature for thermal to quantum nucleation of electron
bubbles  is very low, of the order of 6 mK for $^4$He.

\vspace*{0.5cm}


\end{abstract}

\section{INTRODUCTION}

It has been recognized that liquid helium
is especially well suited for homogeneous cavitation
studies. On the one hand, it can be prepared in a high-purity
state, avoiding heterogeneous cavitation
driven by impurities in the liquid. Besides,
experimental techniques have been developed
\cite{Nis89,Xio91,Lam98,Bal02}
that focus a short burst of ultrasound into a small volume
of bulk liquid, thus preventing cavitation on defects
at the walls of the experimental cell.
On the other hand, helium remains liquid down to zero
temperature ($T$). This  allows to address quantum cavitation,
a phenomenon that may appear at very low 
temperatures.\cite{Lam98,Sat92,Bal95}

Heterogeneous cavitation produced by 
impurities purposely introduced in the liquid is also interesting by
itself,\cite{Cha02} and in a series of recent experiments 
the case of heterogeneous
cavitation caused by electrons (electron bubble explosions)
has been addressed in detail.\cite{Cla96,Cla98,Su98}
Another interesting case of heterogeneous cavitation
in liquid $^4$He thoroughly studied is that caused
by the presence of quantized vortices  acting as
cavitation seeds.\cite{Dal92,Mar94,Bar99}

In this work we attempt a theoretical description of
electron bubble explosions using a $T$-dependent density
functional approach we have employed in a series of studies on
cavitation and nucleation in liquid helium
(see Ref. \onlinecite{Bar02} for a comprehensive review) in conjunction
with a realistic electron-helium (e-He) effective potential. Our
results are in agreement with those obtained in Refs.
\onlinecite{Cla96} and \onlinecite{Cla98} for $^4$He, and
in Ref. \onlinecite{Su98} for $^3$He, covering a wider
temperature range.

This paper is organized as follows. Sections 2 and 3 are mostly 
devoted to $^4$He and to the general formalism.
In Sec. 2 we present the
results obtained using the capillary approximation.
In the case of electron bubbles, this approximation has been
considered realistic
enough to yield semi-quantitative results for critical
pressures and thermal to quantum crossover temperatures,
and constitutes a useful guide to the  results obtained within
density functional (DF) theory. In Sec. 3 we present
the DF plus Hartree  electron-effective potential
approach together with the results obtained for $^4$He 
using this method.
In Sec. 4 we present the results obtained for $^3$He, and
a brief summary is presented in Sec. 5.
A preliminary version of part of this work has been presented
elsewhere.\cite{Pi04}

\section{CAPILLARY MODEL}

In this simple model, the electron is confined in an
impenetrable spherical well potential of radius
$R$. The total energy of the e-He system can be
written as a function of the radius as\cite{Fer57,Kon00}
\begin{equation}
U(R)=\frac{\pi^2 \hbar^2}{2 m_e R^2} + 4 \pi R^2 \sigma
+\frac{4}{3} \pi R^3 P
-\xi \frac{\epsilon -1}{\epsilon} \frac{e^2}{2 R} \,\,\, ,
\label{eq1}
\end{equation}
where $P$ is the pressure applied to the system,
$\sigma$  is the surface tension of the
liquid,  and $\epsilon$ 
is its dielectric constant. The first term is the energy of 
the electron in the ground-state  of an infinite well
potential of radius $R$. For $^4$He the last term can be evaluated taking
$\epsilon = 1.0588$ (Ref. \onlinecite{Ros95}) and $\xi = 1.345$
(Ref. \onlinecite{Kon00}).
Its effect is small and it will not be
considered in the following. 
We have also checked that the effect of including 
a curvature energy term in Eq. (\ref{eq1}) is small.
On the contrary, the effect of the
surface tension on any cavitation process is crucial, and quantitative
results can only be obtained with the use of the correct value of $\sigma$.
In the following, we will take\cite{Roc97} $\sigma=0.272$ K \AA$^{-2}$,
instead of the value $\sigma=0.257$ K \AA$^{-2}$  
given in Ref. \onlinecite{Iin85} that
we and other authors have  used in the past.\cite{Cla96,Cla98,Bar99,Gui92}
We want to mention that the value of Roche
et al.\cite{Roc97} agrees well with that
of Guo  et al.\cite{Guo71} obtained long time ago.

When $P\ge0$, Eq. (\ref{eq1}) has an absolute minimum located
at $R_{min}\,=18.9 \,$ \AA $\,$ at $P=0$.
This configuration corresponds to a stable electron bubble.
When the liquid is depressed below its saturation vapor pressure,
the absolute minimum becomes local, and
$U(R)$ also displays a local maximum. 
The metastability region in the $P-T$ plane extends between
the liquid-vapor coexistence line down to the instability line.
Consequently, metastable bubbles can be formed at positive and
negative pressures as well.
In that region, the electron bubbles are
metastable, and an energy barrier of height
$\Delta U=U(R_{max})-U(R_{min})$ appears which can be
overcome either by
thermal activation above the barrier or by quantum tunnelling
through it. 

\begin{figure}[ht]
\centerline{\includegraphics[width=10cm,height=6.5cm,clip]{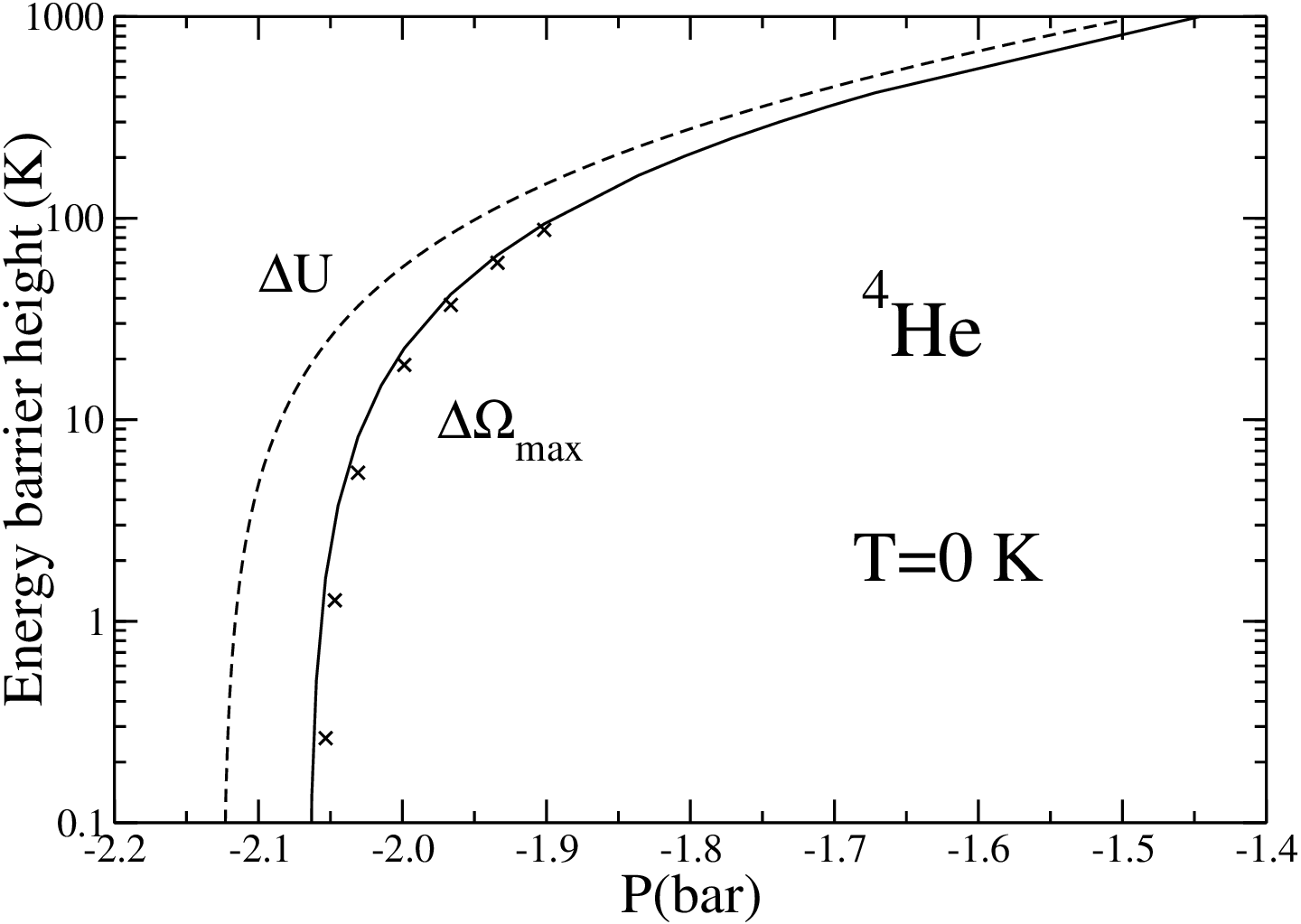} }
\caption[]{
Energy barrier height $\Delta U$ (K) (dashed line)
as a function of $P$ (bar)
of an electron bubble in liquid $^4$He at $T=0$ in the capillary model.
Also shown is the corresponding DF result (solid line).
The crosses along the DF curve have been obtained using a simpler
electron-He interaction, see text.
 }
\label{fig1}
\end{figure}

The height of the energy barrier $\Delta U$
is displayed in Fig. \ref{fig1} as a function of $P$ at $T=0$ K.
It can be seen that if the
negative pressure is large enough, the barrier eventually
disappears and the system becomes globally unstable.
This happens at an instability pressure $P_u$ given by the
expression:

\begin{equation}
P_u=-\frac{16}{5}\,\left(\frac{2 \pi m_e}{5\hbar^2}\right)^{1/4}
\sigma^{5/4}
\label{eq2}
\end{equation}
For the parameters we use, $P_u=-2.12$ bar, and the radius
of the corresponding electron bubble is 
$R_u = 28.2 \,$ \AA.
Had we taken into account the last term in Eq. (\ref{eq1}),
we would have obtained $P_u = -2.24$ bar and
$R_u =  27.1\,$ \AA.
The instability pressure can be compared
to the spinodal pressure at which  pure $^4$He liquid becomes
macroscopically unstable, $P_{sp} = -9.20$ bar.\cite{Gui92}
This value of $P_{sp}$ is consistent with the value obtained by other
authors.\cite{Cla98,Xio89,Bor94}

A necessary condition for the validity of the capillary model is
that the metastable electron bubble has a fairly large radius.
Only in this case one may split the system into a volume 
and a surface region which justifies
the use of Eq. (\ref{eq1}). This happens when
the critical configurations are large
enough, as for example in the case of cavitation occuring near the
liquid-vapor coexistence line.\cite{Xio89,Jez93}
However, in the case of electron bubble explosions the situation
is more complex because
the e-He interaction is strongly repulsive and small changes
in the electron wave function may cause a sizeable effect
on the metastable bubble. 
It is obvious that the bubble configurations used in the
capillary model have little flexibility, and consequently,
the validity of  this approximation should eventually
rely on the comparison with more realistic methods,
as the DF approach. This will be done in Sec. 3.

In most cases, cavitating liquids undergo phase separation before
they reach the stage of global unstability. This proceeds through the
formation  of critical bubbles either by thermal or by quantum
activation. We will show that for electron bubbles, the critical
pressure $P_{cr}$ at which it happens is very close to $P_u$,
a result already obtained in Ref. \onlinecite{Cla98}.
As a general rule, the presence of impurities in the liquid
results in a sizeable decrease of $|P_{cr}|$.
Quantized vortices in  liquid $^4$He play the same role as
impurities, and their presence also decrease $|P_{cr}|$
(Refs. \onlinecite{Dal92,Mar94}) as well as the degree of 
critical supersaturation
in $^3$He-$^4$He liquid mixtures.\cite{Bar99}

Within the capillary approximation, the dynamical evolution of the  
electron bubble can be
parametrized by one  single collective variable, namely
the radius of the bubble. In this case, it is rather simple to
describe quantum and thermal
cavitation regimes on the same footing, continuously passing from one
to the other. This is accomplished by the use of the functional
integral method (FIM)\cite{Col77,Chu92} thoroughly described in 
Ref. \onlinecite{Bar99}. We now recall its essentials.

The nucleation rate $J$ for a thermally activated process,
i.e., the number of critical bubbles formed
in the system per unit time and volume, is
\begin{equation}
J_T = J_{0T} \exp (-\Delta U/ k_B T)    \,\,\, ,
\label{eq3}
\end{equation}
where $k_B$ is the Boltzmann constant.
The prefactor $J_{0T}$ depends on the dynamics of the
process, and it is of the order of the number of cavitation sites
per unit volume (the number of electrons per unit volume, 
$n_e$, in the present case) times an attempt frequency  $\nu_T$.
For simplicity, we take
$\nu_T= k_B T/h $, where $h$ is the Planck constant.

At low enough $T$, thermal activation is no longer possible.
However, cavitation may proceed by quantum tunnelling.
The transition from one regime  to the other is very abrupt, so that a
thermal-to-quantum crossover temperature
$T^*$ may be defined by indicating whether nucleation takes
place thermally $(T > T^*)$ or quantically $(T < T^*)$.
In the limit of zero temperature, cavitation is purely quantal, but
for $T^* > T > 0$ thermally assisted quantum cavitation is the
physical process.

For $T < T^*$ the tunnelling rate is
\begin{equation}
J_Q = J_{0Q} \,\exp(- S^Q/\hbar)     \;  ,
\label{eq4}
\end{equation}
where ${\cal P}= \exp(-S^Q/\hbar)$ is the tunnelling probability,
$S^Q$ is the quantum action, and the prefactor
$J_{0Q}$ is again of the order of the number of nucleation sites per unit
volume times an attempt frequency $\nu_Q$ which can be estimated from
the zero point motion of the system about the metastable equilibrium
position $R_{min}$.

An analytical expression for $T^*$ is obtained which involves
the second derivative of $U(R)$ and the value of the collective mass 
of the bubble $M(R)$ at the maximum of the cavitation barrier,
$R_{max}$
\begin{equation}
k_B T^* = \frac{\hbar}{2\pi}\sqrt{\left. \frac{-1}{M(R_{max})} \frac{d^2
U}{d R^2} \right|_{R_{max}}} \; .
\label{eq5}
\end{equation}
If the motion is irrotational and the fluid incompressible,
the collective mass $M$ depends on $R$ as\cite{Su98,Bar99}
\begin{equation}
M(R)= \frac{4 \pi}{\hbar^2} R^3 m_{He} \rho_b \; ,
\label{eq6}
\end{equation}
where $\rho_b$ is the particle density of the metastable
bulk liquid and $m_{He}$ is the mass of a helium atom.
Eq. (\ref{eq5}) shows that  the value of $T^*$
is determined by {\em  small variations around $R_{max}$}.
An expression for the attempt frequency in the
quantum regime can be also worked out easily:
\begin{equation}
\nu_Q = \frac{1}{2\pi\hbar}\sqrt{\left.\frac{1}{M(R_{min})} \frac{d^2
U}{d R^2} \right|_{R_{min}}} \; .
\label{eq7}
\end{equation}

\begin{figure}[tbh]
\centerline{\includegraphics[width=10cm,height=6.5cm,clip]{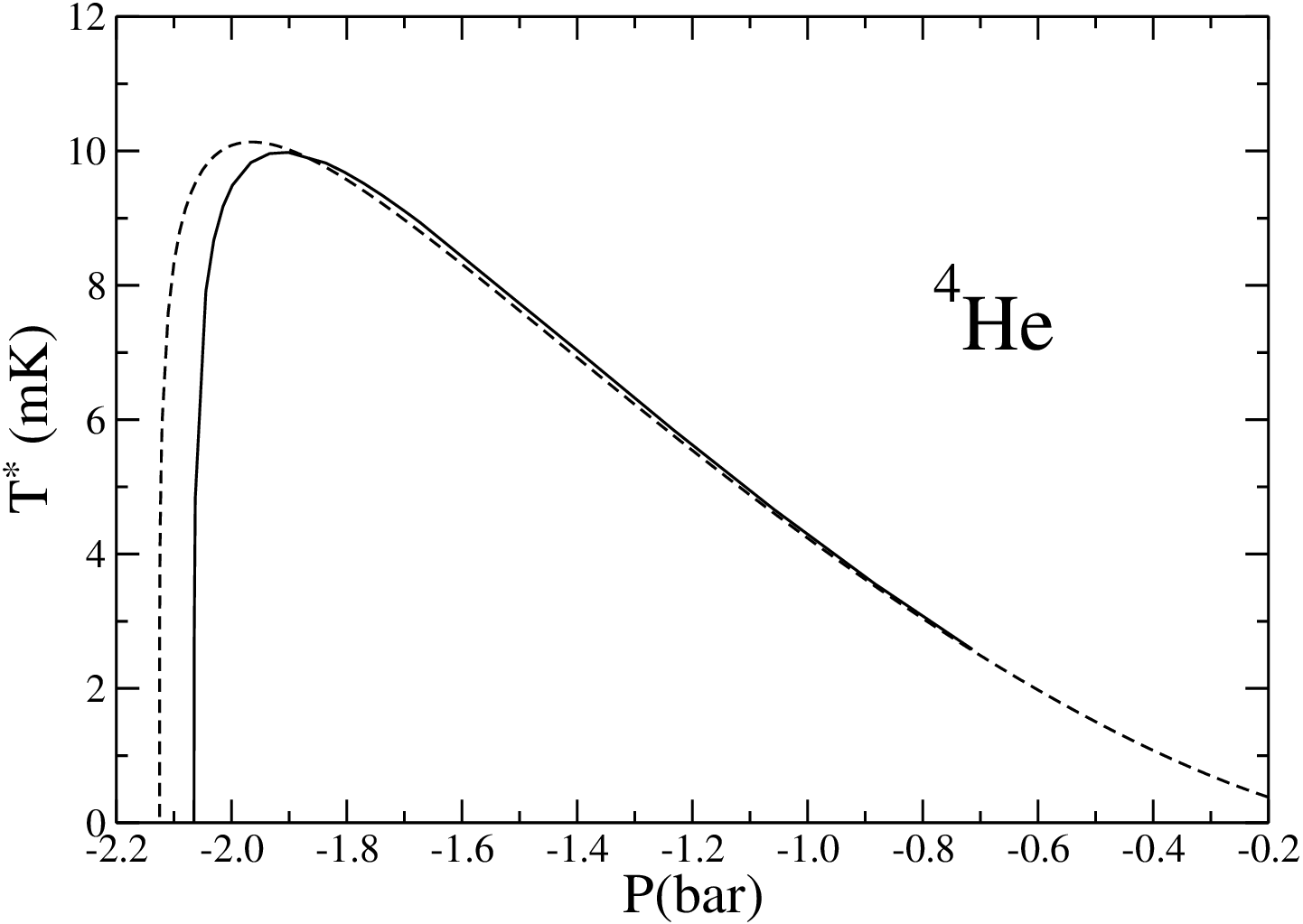} }
\caption[]{
Crossover temperature $T^*$ (mK) as a function of $P$ (bar)
of an electron bubble in liquid $^4$He in the capillary model
(dashed line) and in the DF approach (solid line).
 }
\label{fig2}
\end{figure}

Using Eq. (\ref{eq5}) we have determined $T^*(P)$ and
the results are shown in Fig. \ref{fig2}. This
figure shows that in the capillary approximation,
 irrespective of the pressure, above $T \sim 10$ mK
the cavitation process is thermal and not quantal.
To determine which
$T^*$ corresponds to the actual experimental conditions, one has to
look for the intersection of the curve $T^*(P)$ with the line that
results imposing that  critical
bubbles nucleate with appreciable probability inside the experimental
volume during the experimental time:
\begin{equation}
1= t_{exp} V_{exp}\, J_{0Q} e^{-S^Q_{min}/\hbar}=
t_{exp} V_{exp}\, J_{0T} e^{-\Delta U(P)/k_B T^*} \; ,
\label{eq8}
\end{equation}
where $t_{exp}$ and $V_{exp}$ are the experimental time and volume,
respectively, and $J_{0Q}= n_e \nu_Q$. Taking\cite{Cla98}
$t_{exp} \sim 10^{-5}$ s, $V_{exp} \sim 10^{-5}$ cm$^3$,
$n_e= 10^6$ cm$^{-3}$, and the value $\nu_Q \sim 10^9$ s$^{-1}$ obtained
from Eq. (\ref{eq7}), one gets $T^* = 4.7$ mK at a pressure
slightly above $P_u$. 
At
$T^* = 4.7$ mK one has $S^Q = 11.5\, \hbar$, so that the use of
the formalism of Ref. \onlinecite{Bar99} is well 
justified.\cite{Chu92}
Had we used the WKB approximation\cite{Su98a} to estimate $T^*$, we would have
obtained similar results. In particular, the maximum
of $T^*_{WKB}$ is  $9.6$ mK.
We will see in Sec. 3 how these results change in the DF approach.

For a given temperature above $T^*$, the solution to Eq. (\ref{eq8})
with $J_T$ instead of $J_Q$ yields the pressure
at which critical electron bubbles are created with
sizeable probability. For the mentioned experimental parameters,
it turns out that in the $T \sim 1$ K regime this happens
when $\Delta U \sim 11.5$ K. Inspection of
Fig. \ref{fig1} readily shows that  this pressure is
very close to $P_u$. Consequently, $P_u$ is the key quantity
for electron bubble explosions.\cite{Cla98}

We would like to recall why
the capillary model is expected to be fairly realistic for electron
bubble explosions whereas it is completely unrealistic  to
address homogeneous cavitation in liquid helium. Dropping the
electron contributions to Eq. (\ref{eq1}), it is easy to see that
for negative pressures
$\Delta U= 16 \pi \sigma^3/3 |P|^2$ at
$R_{max} = 2 \sigma/|P|$. This means that the
barrier height is not zero at the spinodal pressure, which is very
unphysical. This drawback renders useless the capillary
model for homogeneous cavitation in pure liquid helium at low $T$
because it has been  experimentally established that cavitation occurs
near the spinodal region.\cite{Lam98,Bal02,Cau01}
On the other hand, near $R_{min}$ the model yields  empty
bubbles of very small radius, which is also unphysical.
As a consequence, any  dynamical approach based on the  use of these
bubble configurations is rather dubious.\cite{Nak95} The model can only work
well for large bubbles, as for example near the saturation curve. In
these situations, it has been shown that it yields realistic homogeneous cavitation
pressures,\cite{Nis89,Bal02,Sin82} 
and the use of Eq. (\ref{eq5}) to obtain
$T^*$ yields values in good agreement with the DF ones.\cite{Bar02,Gui96} 
The situation is quite similar to that found in
the description of supersaturated $^3$He-$^4$He liquid mixtures, see Refs.
\onlinecite{Bar99} and \onlinecite{Tan02} for a thorough discussion.

\section{DENSITY-FUNCTIONAL APPROACH TO ELECTRON BUBBLE
EXPLOSIONS}

Density functional theory has been applied in the past to cavitation
in classical liquids (see Ref. \onlinecite{Oxt92} and
Refs. therein).
Since the pioneering work of Xiong and Maris,\cite{Xio89}
it has proven to be the most successful approach in
addressing cavitation in liquid helium so far.\cite{Bar02} It
incorporates in a self-consistent way
the equation of state of bulk liquid
and surface tension of the liquid-gas interface as a function of
temperature, which are key ingredients of any cavitation model.
It does not impose a priori the density profile of the
critical cavity,
allowing for a flexible description of the process from the
saturation line down to the spinodal line. 
Moreover,
within DF theory one avoids to split the system into a bulk and a surface
region, and the use of macroscopic concepts such as surface tension
and pressure at a nanoscopic scale.
However, it is a continuous, not an atomic description of the system.
In spite of this, it has been found to well describe situations in which the
atomic scale is relevant, such as quantized vortices, or the presence of
strongly attractive atomic or molecular impurities (see for instance
Ref. \onlinecite{Anc03} and Refs. therein).

In the frame of DF theory,
the properties of an electron bubble approaching the surface of liquid
$^4$He have been studied by Ancilotto and Toigo.\cite{Anc94} They have used 
the so-called Orsay-Paris  zero temperature finite-range DF,\cite{Dup90} 
and the pseudopotential
proposed in Ref. \onlinecite{Kes65} as e-He interaction.
The method chosen by Classen et al.\cite{Cla98} is a simplification of
that of Ref. \onlinecite{Anc94} in two respects. Firstly, they have used
a  zero-range DF to describe $^4$He, which seems justified
in view of the slowly varying helium densities even in the presence of
excess electrons in the liquid (the situation is completely
different for strongly attractive atomic or molecular impurities, 
see for example Ref. \onlinecite{Dal94}).
However, they have included some thermal effects in the DF, whereas
the approach of Ancilotto and Toigo is at zero temperature.\cite{Anc94}
Secondly, the pseudopotential has been replaced by a contact e-He
interaction
whose intensity has been adjusted so as to reproduce the total energy of
an excess electron in bulk helium,\cite{Som64} which is about 1 eV for a
particle density of $0.0218$ \AA$^{-3}$ (saturation density).

Our starting point is a finite temperature zero-range DF that
reproduces thermal properties of liquid $^4$He such as
the experimental isotherms
and the $^4$He liquid-gas coexistence line up to $T = 4.5$ K, and the
$T$ dependence of the surface tension of the liquid free surface.\cite{Gui92}
This DF has been successfully used to address homogeneous
cavitation in liquid helium from $T \sim 0$ K up to temperatures close to the 
critical one.\cite{Bal02} In the quantum cavitation regime, it has also yielded
results in good agreement with experiment\cite{Lam98} and with
other theoretical approaches.\cite{Mar95} We
have taken the Hartree-type e-He effective potential
derived by Cheng et al.\cite{Che94} (see also Ref. \onlinecite{Ros95})
as e-He interaction.
This allows us to write the free energy of the system as
a functional of the $^4$He particle
density $\rho$, the excess electron wave function $\Psi$, and $T$:
\begin{equation}
F[\rho, \Psi, T] = \int d\vec{r}\, f(\rho, T)+
\frac{\hbar^2}{2\,m_e} \int d\vec{r}\, |\nabla \Psi(\vec{r}\,)|^2
+ \int d\vec{r}\, |\Psi(\vec{r}\,)|^2 V(\rho)  \; ,
\label{eq9}
\end{equation}
where $f(\rho, T)$ is the $^4$He free energy density per unit volume 
written as
\begin{equation}
f(\rho,T) = f_{vol}(\rho,T) + \beta \frac{(\nabla \rho)^2}{\rho}
+ \xi (\nabla \rho)^2  \; .
\label{eq10}
\end{equation}
In this expression,
$f_{vol}(\rho,T)$ consists of  the well-known free energy density 
of a Bose gas, plus phenomenological density dependent terms that take into
account the effective interaction of helium atoms in the bulk liquid.\cite{Gui92}
The parameters of these terms and those of the density gradient terms
in Eq. (\ref{eq10}) have been adjusted so as to reproduce physical quantities
such as the equation of state of the bulk liquid and the
surface tension of the liquid free surface. We have slightly modified
the original value\cite{Gui92} of the parameter $\xi$ in Eq. (\ref{eq10})
to exactly reproduce the surface tension of liquid $^4$He
(Ref. \onlinecite{Roc97}), taking $\xi=2330$ K \AA$^5$.
The $\beta$-term is a kinetic energy term; at $T=0$ the system is
described as a Bose condensate and for this reason the kinetic
energy arises only from the inhomogeneity of the 
density.\cite{Str87}
For inhomogeneous systems, this term is essential to have
densities well behaved everywhere, and in the case of
$^4$He droplets its inclusion in the functional yields
densities that smoothly -exponentially- go from
the  bulk liquid down to zero.
We have taken\cite{Str87} $\beta= (\hbar^2/2m_4)/4$.

In bulk helium, knowledge  of $f(\rho, T)$
enables solution of the phase equilibrium equations and to determine
the spinodal line. It also yields an
equation of state in the negative pressure regime, inaccessible to the
experimental determination, through the thermodynamic relationship
$P = - f_{vol}(\rho,T) + \mu \rho$, where $\mu$ is the $^4$He chemical
potential.

The e-He interaction $V(\rho)$ is written as a function of the local
helium density\cite{Che94}
\begin{equation}
V(\rho) = \frac{\hbar^2 k^2_0}{2\,m_e}
+\frac{2 \pi \hbar^2}{m_e} \rho \,a_{\alpha}
- 2 \pi \alpha e^2 \left(\frac{4 \pi}{3}\right)^{1/3} \rho^{4/3} \; ,
\label{eq11}
\end{equation}
where $\alpha = 0.208\,$ \AA$^3$ is the static polarizability of
a $^4$He atom, and $k_0$ is determined from the helium local Wigner-Seitz
radius $r_s = (3/4 \pi \rho)^{1/3}$ by solving the transcendent equation
\begin{equation}
\tan[k_0(r_s-a_c)] = k_0\,r_s \; ,
\label{eq12}
\end{equation}
with $a_c$ and $a_{\alpha}$ being the scattering lengths arising from a
hard-core and from a polarization potential. 
We have taken\cite{Che94}
$a_{\alpha} = -0.06\,$ \AA, $a_c = 0.68\,$ \AA.

The application of DF theory to the cavitation problem proceeds 
as follows. For given $P$ and $T$ values
one first determines the metastable and unstable cavities
that would correspond to the local minimum and
maximum configurations in the capillary model (actually, in
the multidimensional space spanned by the more flexible DF configurations,
the latter is no longer a local maximum but a saddle point).
This is achieved by solving the Euler-Lagrange equations which result
from the variation of the constrained grand potential density 
$\tilde{\omega}(\rho, \Psi, T) = \omega(\rho, \Psi, T)
- \varepsilon |\Psi|^2 $, where the grand potential density
$\omega(\rho, \Psi, T)$ is defined from Eq. (\ref{eq9}) as
\begin{equation}
\omega(\rho, \Psi, T) =  f(\rho, T)+
\frac{\hbar^2}{2\,m_e} |\nabla \Psi|^2
+ |\Psi|^2 V(\rho) - \mu \rho \; .
\label{eq13}
\end{equation}
It yields
\begin{equation}
\frac{\delta f}{\delta \rho} +|\Psi|^2\, \frac{\partial V}{\partial \rho}
= \mu
\label{eq14}
\end{equation}
\begin{equation}
-\frac{\hbar^2}{2\, m_e}\Delta \Psi + V(\rho) \Psi  = \varepsilon \Psi
  \; ,
\label{eq15}
\end{equation}
where $\varepsilon$ is the lowest eigenvalue of the Schr\"odinger
 equation obeyed by the electron.
These equations are solved assuming spherical symmetry, imposing for $\rho$
the physical conditions that $\rho'(0) = 0$ and
$\rho(r \rightarrow \infty) = \rho_b$, where $\rho_b$ is the density
of the metastable bulk liquid, and that the electron is in the $1s$ state.
Fixing $\rho_b$ and $T$ amounts to fix  $P$ and $T$, as the pressure can
be obtained from the bulk equation of state
$P = - f_{vol}(\rho_b,T) + \mu \rho_b$, as well as $\mu$. Thus,
$\mu= \partial f_{vol}(\rho, T)/\partial \rho |_T$ is known
in advance, whereas $\varepsilon$ is not and has to be determined  
from Eq. (\ref{eq15}).

\begin{figure}[h]
\centerline{\includegraphics[width=8cm,height=11cm,clip]{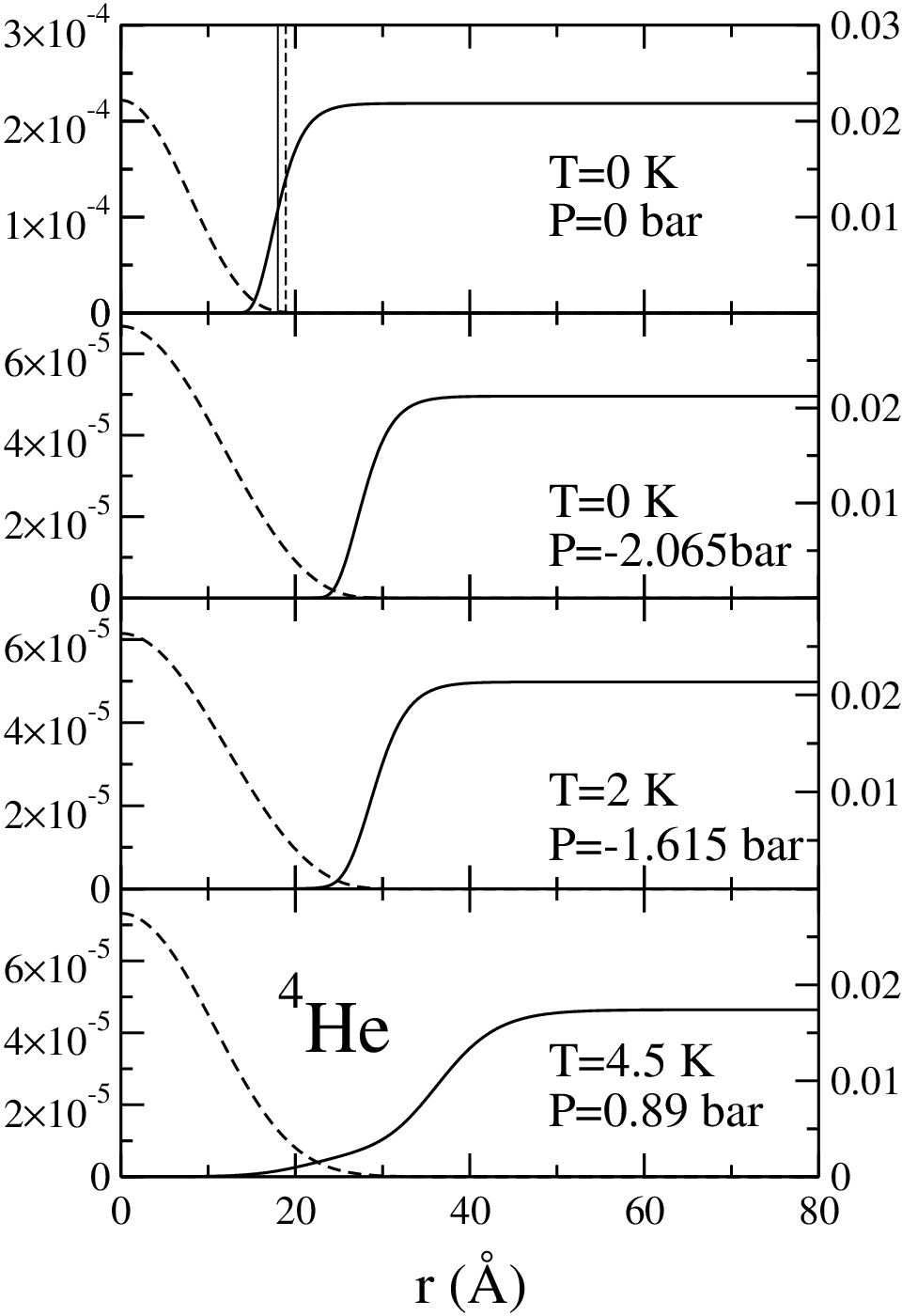} }
\caption[]{
$^4$He density profiles in \AA$^{-3}$ (solid lines, right scale)
and excess electron squared wave functions $|\Psi|^2$ in
\AA$^{-3}$  (dashed lines, left scale),
as a function of radial distance $r$ (\AA).
The top panel shows the stable bubble at $T=0$ K,
$P=0$ bar. The other three panels show
the near-to-unstable electron bubble for several $(P,T)$
values. In the upper panel, the vertical thin dashed line indicates
the radius of the capillary model bubble ($R_{min}=18.9$ \AA) , 
and the vertical thin solid
line indicates the radius at which the helium density equals $\rho_b/2$
[$R(\rho_b/2)= 18.0 $ \AA, with $\rho_b = 0.0218$ \AA$^{-3}$].
 }
\label{fig3}
\end{figure}
\newpage
We have solved Eqs. (\ref{eq14}) and (\ref{eq15})
using a multidimensional Newton-Raphson method\cite{Pre92}
after having discretized them
using $n$-point formulas for the $r$ derivatives. 
We have used $n=13$ formulas, but
comparable results have been obtained using $n=7$ and 9 
formulas.\cite{Bic41}
A fine mesh of step $\Delta r = 0.1\, $\AA$\,$
has been employed, and the equations have been integrated up to
$R_{\infty} = 150\,$ \AA $\,$ to make sure that the asymptotic bulk
liquid has been reached. 
The multidimensional Newton-Raphson method has been applied until 
the {\em local chemical potential} -left hand side of 
Eq. (\ref{eq14})- does not differ
substantially from $\mu$. We have checked that, for every
$r$ value, both coincide up to at least the sixth decimal figure. 
This is a crucial test on the accuracy of our
method. We have thus
achieved a {\it fully variational solution} of the Euler-Lagrange
 problem embodied
in Eqs. (\ref{eq14}) and (\ref{eq15}), valid from $r = 0$ up to $R_{\infty}$.

\begin{figure}[bth]
\centerline{\includegraphics[width=10cm,height=6.5cm,clip]{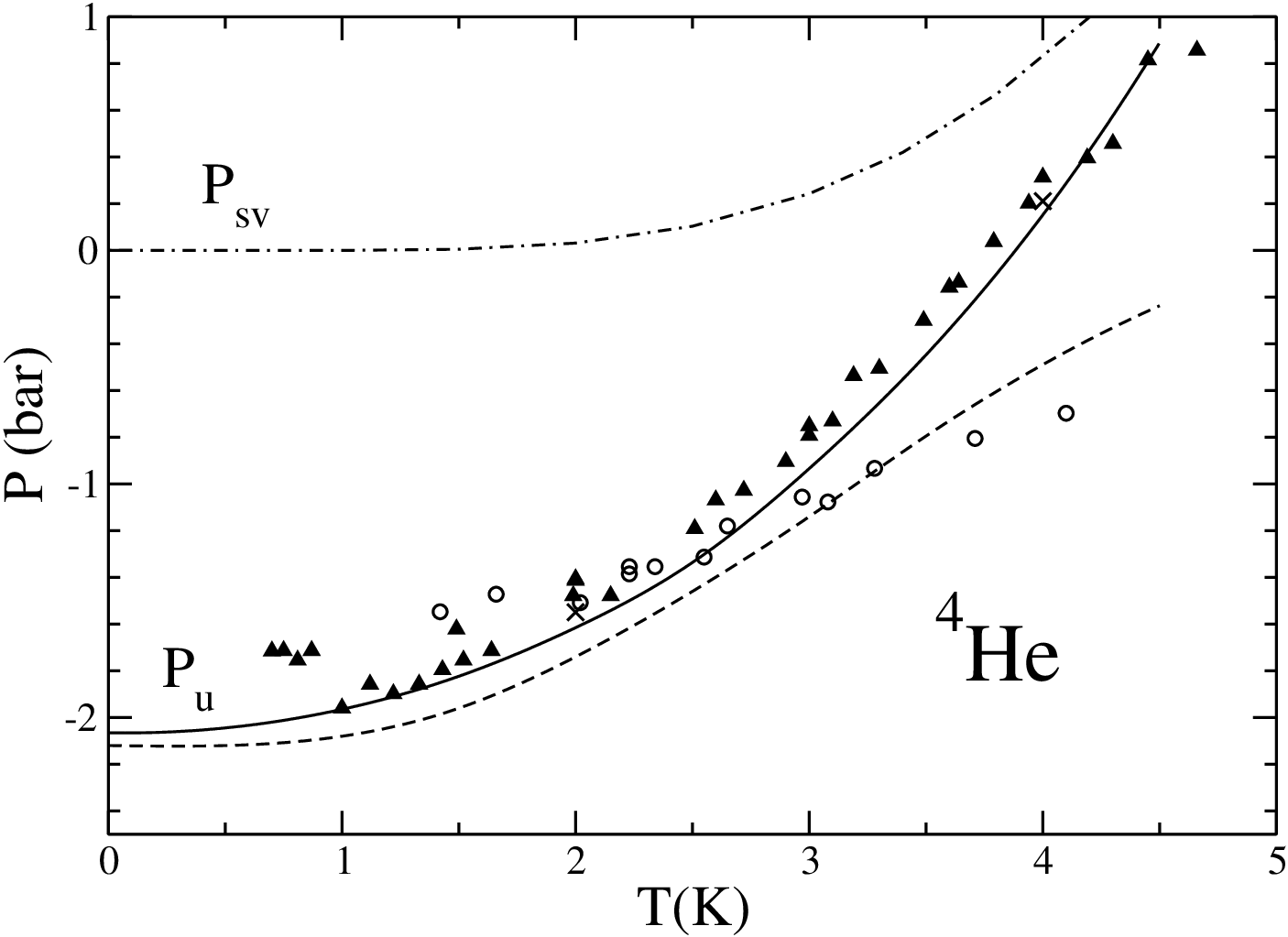} }
\caption[]{
Instability pressure $P_u$ as a
function of $T$ for $^4$He. Circles and triangles are experimental data from
Refs. \onlinecite{Cla96} and \onlinecite{Cla98}, respectively. The
results of the capillary model are represented by a dashed line,
and the DF results by a solid line. 
The dash-dotted line represents
the experimental saturation  vapor pressure $P_{sv}$ line.
The crosses at $T=2$ and 4 K indicate the critical pressures $P_{cr}$.
}
\label{fig4}
\end{figure}

We represent in Fig. \ref{fig3} several $^4$He density profiles
and excess electron squared wave functions $|\Psi|^2$.
The top panel shows the stable bubble at $T=0$ K,
$P=0$ bar. The other three panels
display the near-to-unstable electron bubble for several $(P,T)$
values. For a given $T$, they have been obtained decreasing $\rho_b$,
i.e. $P$, until Eqs. (\ref{eq14}) and (\ref{eq15}) have no solution.
The smaller $P$ value defines $P_u$.
The $^4$He instability pressure $P_u$ is shown in Fig.
\ref{fig4} as a function of $T$.
This figure shows that the lowest pressure the system may reach
before becoming macroscopically unstable is $P_u= -2.07$ bar, which
is the value corresponding to $T=0$ K.

\begin{figure}[hbt]
\centerline{\includegraphics[width=8cm,height=11cm,clip]{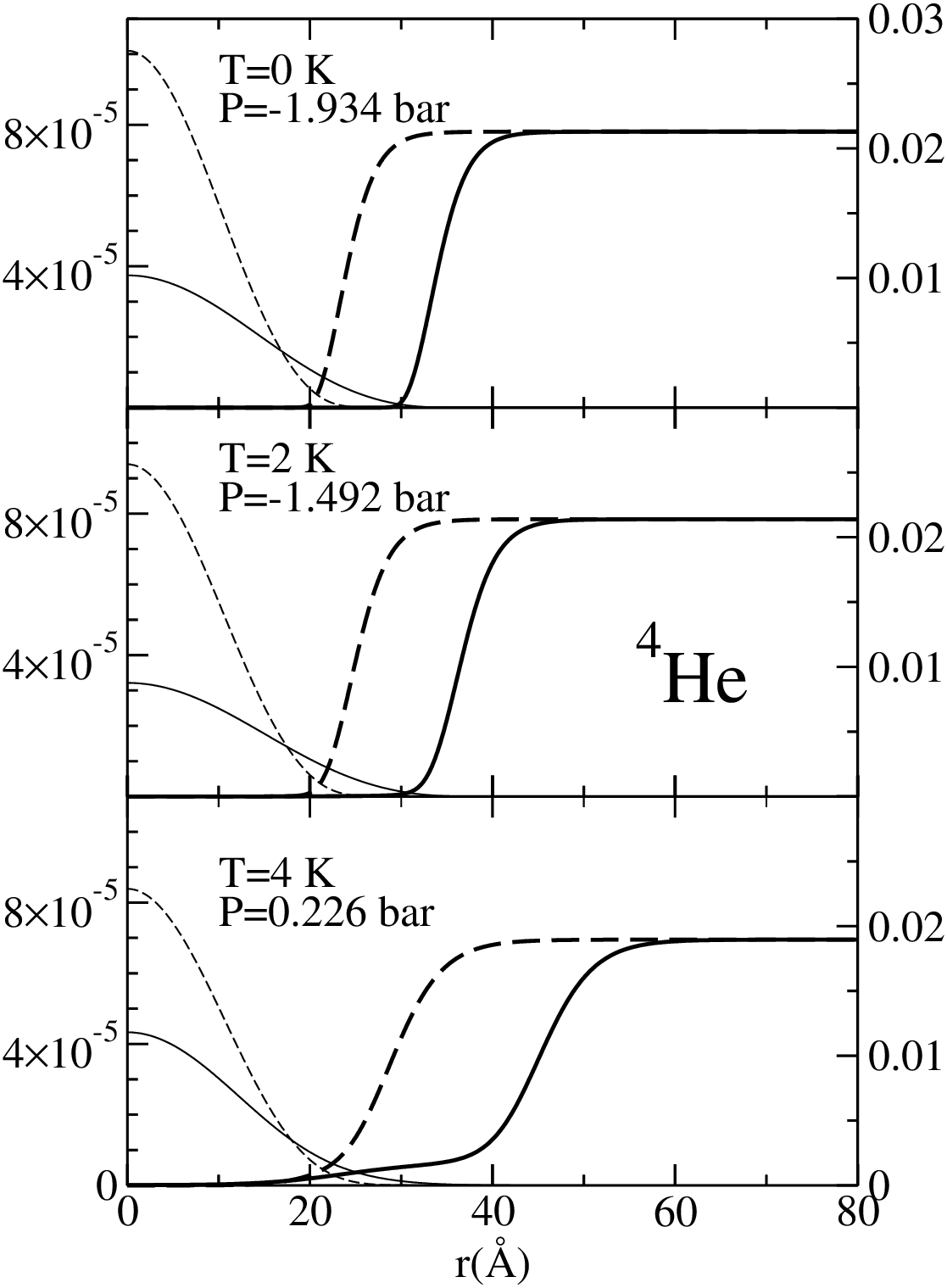} }
\caption[]{
$^4$He density profiles in \AA$^{-3}$ (thick lines, right scale)
and excess electron squared wave functions $|\Psi|^2$ in \AA$^{-3}$
(thin lines, left scale), as a function of radial distance $r$ (\AA)
for the metastable and saddle electron bubble at different
$(P,T)$ values. Dashed lines correspond to the metastable configuration,
and solid lines to the saddle configuration. From top to bottom, the
$\Delta\Omega_{max}$  values are 65.3, 80.2, and 88.1 K,
respectively. 
 }
\label{fig5}
\end{figure}

As we have indicated, in the metastability region the 
Euler-Lagrange equations have
two different solutions for given $\rho_b$ and $T$ -i.e., $P$ and $T$-
values, one corresponding to the metastable configuration and another
corresponding to the saddle configuration. A similar situation is found
in the case of cavitation in the presence of vortices.\cite{Jez97}
Actually, it is the search of the saddle configuration that constitutes
a numerical challenge. The reason is that, due to the strongly repulsive
e-He interaction, fairly small changes in the electron configuration
induce changes in the helium configuration hard to handle numerically,
so that in the course of the numerical procedure, the system has a
strong tendency to jump from the saddle to the metastable solution. We
have represented in Fig. \ref{fig5} several $^4$He density profiles and
excess electron squared wave functions $|\Psi|^2$ corresponding to the
metastable and saddle electron bubbles for $T=0$, 2 and 4 K, and a value
of $P$ close to $P_u$.

The cavitation barrier height $\Delta \Omega_{max}$ is obtained by
subtracting the grand potential of the saddle bubble
$\omega(\rho_s,\Psi_s,T)$  from that of the metastable
bubble $\omega(\rho_m,\Psi_m,T)$:
\begin{equation}
\Delta \Omega_{max} =
\int d\vec{r} \,[ \omega(\rho_s,\Psi_s,T) - \omega(\rho_m,\Psi_m,T)]
\; .
\label{eq16}
\end{equation}

\begin{figure}[tbh]
\centerline{\includegraphics[width=10cm,height=6.5cm,clip]{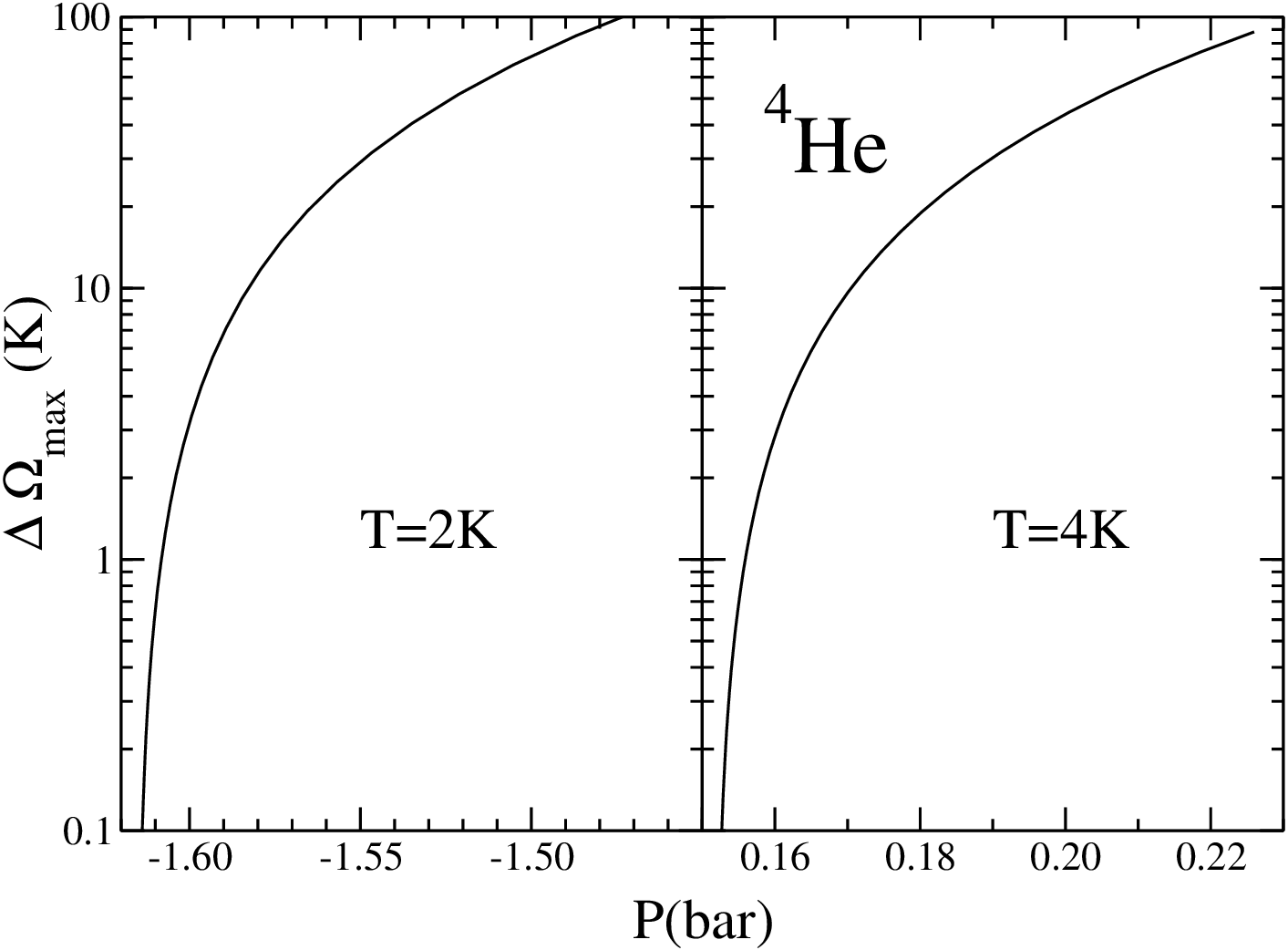} }
\caption[]{
DF energy barrier height $\Delta \Omega_{max}$ (K) for $^4$He
as a function of $P$ (bar) for  $T=2$ K (left panel)
and $T=4$ K (right panel).
 }
\label{fig6}
\end{figure}

Since both configurations go asymptotically to the same $\rho_b$
value, Eq. (\ref{eq16}) gives $\Delta \Omega_{max}$ as a function of
$P$ and $T$ via the equation of state of bulk liquid helium.
$\Delta \Omega_{max}$ is shown in Fig. \ref{fig6}
as a function of $P$ for $T=2$ and 4 K, and in Fig. \ref{fig1} for $T=0$ K.
As indicated, $\Delta \Omega_{max}$ becomes negligible when the system
approaches the unstability pressure. This constitutes a suplementary test on the
correctness of the near-to-unstable configurations we have found by
decreasing $\rho_b$ at fixed $T$. Otherwise, 
$\Delta \Omega_{max}$ would  not be negligible for this
configuration.

\newpage
Once $\Delta \Omega_{max}(P,T)$ has been determined,
it can be used to obtain the critical pressure $P_{cr}$ at which 
critical bubbles nucleate at an appreciable rate
by solving an equation similar to Eq. (\ref{eq8}):
\begin{equation}
1= t_{exp} V_{exp}\, J_{0T} e^{-\Delta \Omega_{max}(P,T)/k_B T} 
\label{eq17}
\end{equation}
taking $J_{0T}= \nu_T \, n_e$. This yields
$P_{cr}=-1.55$ bar at $T=2$ K, and
$P_{cr}=0.211$ bar at $T=4$ K. These pressures
are slightly above the corresponding $P_u$ values which are, respectively,
$-1.62$ and $0.151$ bar.

We thus see from Fig. \ref{fig4} that our results are fully compatible 
with the available
experimental data --the calculated $P_u$ should be a lower bound to the
cavitation pressure--
if the experimental results of Ref. \onlinecite{Cla96} are ruled out,
and those carried out more recently by the same group\cite{Cla98} 
have error bars similar to those found
in the case of pure liquid $^4$He,\cite{Cau01} as well as in the
case of electron bubble cavitation in $^3$He.\cite{Su98}

The capillary model 
yields instability pressures lower than those obtained
within DF theory, especially at high temperatures. The discrepancy 
arises even if one takes into account, as we did, the
$T$-dependence of the surface tension.
The origin of the discrepancy is the e-He
strongly repulsive interation acting on the helium density tail
that penetrates inside the cavity as $T$ increases:
the electron has to `push' not only the bulk surface, which for
DF configurations is at the radius where the helium
density equals $\rho_b/2$, but also the part of the helium density
that is spread inside the bubble, especially at high 
temperatures. This `pushing' produces a steeper bubble density
profile, and not only an actual displacement of its
surface. For this reason, the radius of the bubbles are smaller in
the DF than in the simple capillary approach
(see Figs. \ref{fig3} and \ref{fig8}). Part of the difference
is removed if one allows for penetration of the electron wave
function into the liquid using a finite height barrier for the potential
that confines the electron within the bubble. This would
diminish the bubble radius, as it lowers the zero-point energy
of the electron. However, the final effect is that it increases $|P_u|$ 
and the agreement with the DF result worsens. \cite{Cla98}

Our results for $P_u$ are  slightly below
those of Ref. \onlinecite{Cla98} (compare for instance
our value of -2.07 bar at $T=0$ K with their value of -1.92 bar).
This difference is essentially due to the different value of the
surface tensions used to build the DF used in our work and in theirs.
Indeed, the pressure we obtain if the DF
is adjusted to reproduce the surface tension of Ref. \onlinecite{Iin85}
is $P_u=-1.93$ bar, in excellent agreement with the value found in 
Ref. \onlinecite{Cla98}. Moreover,
the ratio 2.07/1.93 compares very well with that obtained
from Eq. (\ref{eq2}) if ones takes $\sigma = 0.272$ K \AA$^{-2}$
in one case, and $\sigma = 0.257$ K \AA$^{-2}$ in the other.
We recall that in  Ref. \onlinecite{Cla98} the kinetic $\beta$-term 
in  Eq. (\ref{eq10}) has been neglected, which
makes the helium density to be  zero at the origin, and that
these authors have made the helium density strictly zero inside a
sphere around the
excess electron, whereas in our case the helium density is defined
everywhere. These differences do not seem to play any role.

We have also
obtained $\Delta \Omega_{max}$ at $T=0$ K using the contact e-He
interaction used in Ref. \onlinecite{Cla98}.
The results, indicated by crosses in Fig. \ref{fig1}, indicate
that the contact interaction
sensibly yields the same barrier heights as the Hartree-type interaction.

We have employed the DF theory to obtain the crossover
temperature $T^*$.\cite{Gui97} To this end, one has to
obtain the frequency of the small amplitude oscillations 
around the saddle configuration 
in the inverted barrier potential well.
From the capillary model results we expect that $T^*$
is very small, so that thermal effects on $\rho_s$ can be
neglected. In view of the fairly large size of the saddle
bubble, the oscillation frequency
can be obtained as follows.\cite{Gui96,Gui95} After
determining the saddle configuration $\rho_s(r)$, we define a
continuous set of densities by a rigid translation of $\rho_s(r)$:
\begin{equation}
\rho_{\delta}(r) \equiv \left\{ \begin{array}{ll}
   \rho_s(r=0) & {\rm if} \; r \le \delta \\
   \rho_s(r-\delta ) & {\rm if} \; r \ge \delta \\
\end{array}
\right.
\label{eq18}
\end{equation}
The variable $\delta$ roughly represents the displacement of the
surface of the saddle configuration with respect to its stationary value.
Varying $\delta$, all physically relevant configurations are generated.
Eq. (\ref{eq18}) implies that
the surface diffuseness of the bubble is kept frozen during the
displacement.

The barrier is then  obtained as a function of $\delta$:
\begin{equation}
\Delta \Omega_{\delta} =
\int d\vec{r} \,[ \omega(\rho_{\delta},\Psi_{\delta},T=0) -
\omega(\rho_m,\Psi_m,T=0)] \; .
\label{eq19}
\end{equation}
Within this model, $\delta$ is the only collective
variable describing the bubble oscillation, and all the time-dependence
is in $\delta(t)$. The kinetic energy associated with the
oscillation is
\begin{equation}
E_{kin}=\frac{m_{He}}{2} \int {\rm d}{\vec r} \,\rho({\vec r},t)\, {\vec u}^{\,2}
({\vec r}, t)   \; ,
\label{eq20}
\end{equation}
where ${\vec u}({\vec r},t)$ is the velocity field which can be
formally obtained from the continuity equation
\begin{equation}
\frac{\partial \rho}{\partial t} + {\vec \nabla} (\rho \,{\vec u}\,) = 0   \; .
\label{eq21}
\end{equation}
It yields:
\begin{equation}
u(r,t)=-\frac{1}{r^2\rho(r,t)} \int_{0}^{r} s^2\,
\dot{\rho}(s,t)\, {\rm d}s     \, .
\label{eq22}
\end{equation}
By construction,
\begin{equation}
\rho(r,t)=\rho_s(r-\delta(t)) \,.
\label{eq23}
\end{equation}
Thus
\begin{equation}
\dot{\rho}(r,t) = -\rho_\delta^\prime (r)\dot{\delta} 
\label{eq24}
\end{equation}
and
\begin{equation}
u(r,t)=\frac{\dot{\delta}}{r^2\rho_\delta(r)}\bigg[r^2\rho_\delta(r)
-2\int_{0}^{r}s\rho_\delta(s){\rm d}s\bigg] \,.
\label{eq25}
\end{equation}
Defining the mass parameter M$(\delta)$ as
\begin{equation}
E_{kin} \equiv \frac{\hbar^2}{2} M(\delta) \dot{\delta}^2
\label{eq26}
\end{equation}
we get
\begin{equation}
M(\delta)=\frac{4\pi m_{He}}{\hbar^2}\int_{0}^{\infty}\frac{{\rm d}r}
{r^2\rho_\delta(r)}\bigg[r^2\rho_\delta(r)-2\int_{0}^{r} {\rm d}s\,
s\rho_\delta(s)\bigg]^2  \; .
\label{eq27}
\end{equation}
Proceeding as in the capillary case one obtains
\begin{equation}
k_B T^* = \frac{\hbar}{2\pi}\sqrt{-\left.\frac{\partial^2
\Delta\Omega}{\partial \delta^2}\bigg/ M(\delta) \right|_{\delta=0}} 
\;\;\; ,
\label{eq28}
\end{equation}
which is the generalization of Eq. (\ref{eq5}) to the case of diffuse
density profiles. Using this equation we have obtained the $T^*(P)$
curve shown in Fig. \ref{fig2}, and proceeding as in the capillary
model, we have determined a crossover temperature of 6.0 mK.

\section{ELECTRON BUBBLE EXPLOSIONS IN LIQUID $^3$He}

\begin{figure}[bth]
\centerline{\includegraphics[width=10cm,height=6.5cm,clip]{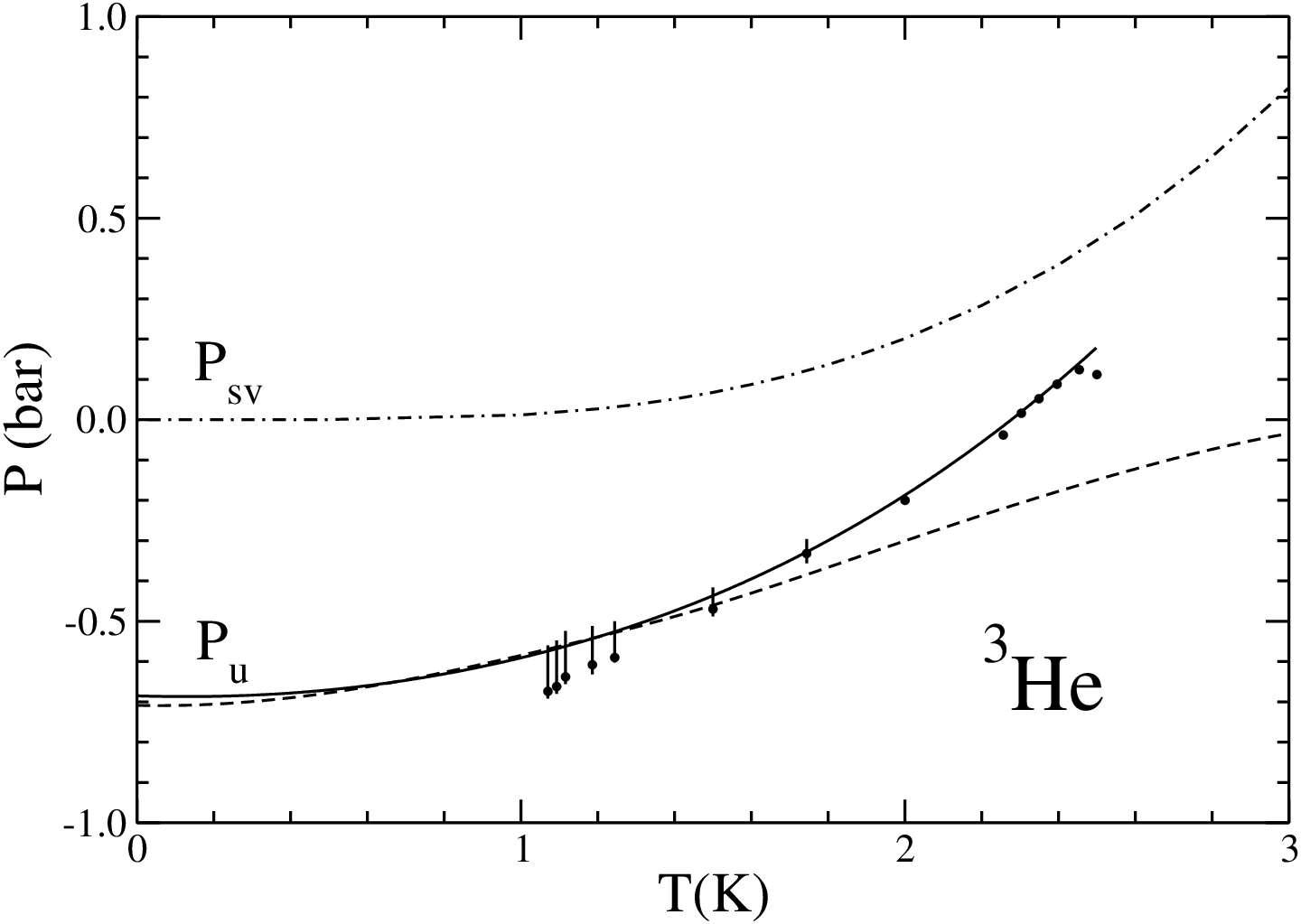} }
\caption[]{
$^3$He instability pressure $P_u$ as a
function of $T$ for $^3$He. Dots are experimental data from
Ref. \onlinecite{Su98}. The dash-dotted line represents
the experimental saturation  vapor pressure $P_{sv}$ line.
The results of the capillary model are represented by a dashed line,
and the DF results by a solid line. 
}
\label{fig7}
\end{figure}

We have also studied the explosion of electron bubbles in
the case of liquid $^3$He. The capillary model of
Sec. 2 can be straightfowardly applied to  this isotope  using the
appropriate values of the surface tension $\sigma = 0.113$ K \AA$^{-2}$
(Ref. \onlinecite{Iin85b}, where one may also find the values of
$\sigma(T)$ we have used to obtain the capillary model results
we show in Fig. \ref{fig7}) and of the
dielectric constant $\epsilon = 1.0428$ (Ref. \onlinecite{Ros95b})
As in Sec. 2, we have neglected  this term in the calculations
because of its smallness.

At zero pressure and temperature, the capillary model yields 
an electron bubble of radius $R_{min} = 23.5 \,$ \AA, larger than
for $^4$He because of the smaller $^3$He surface tension. For the
same reason, the $^3$He instability pressure $P_u = -0.71 $ bar 
is smaller in absolute value [see Eq. (\ref{eq2})]. This
pressure is atteined for an electron bubble of radius
$R_u = 35.2 \,$ \AA.

The application of DF theory to describe electron bubble explosions in
$^3$He proceeds as indicated in Sec. 3. We have used the functional 
proposed in Ref. \onlinecite{Bar90} which we have employed in the 
past\cite{Jez93} to describe homogeneous cavitation in
liquid $^3$He, and the e-He interaction given in
Eq. (\ref{eq11}) with the parameters corresponding to $^3$He, namely
$\alpha = 0.206\,$ \AA$^3$, and same values for $a_{\alpha}$  and
$a_c$.

The $^3$He instability pressure $P_u$ is shown in Fig.
\ref{fig7} as a function of $T$.
This figure shows that the lowest pressure the system may reach
before becoming macroscopically unstable is $P_u=- 0.69$ bar, which
is again the value corresponding to $T=0$ K. For the reason
indicated before, we have not considered necessary to
calculate the critical pressures in the case of $^3$He.
It can be seen that our values of $P_u$ are somewhat above the
experimental values of $P_{cr}$ obtained in Ref. \onlinecite{Su98},
especially at $T \sim 1$ K. As in the $^4$He case, the capillary model
fails as soon as thermal effects start being sizeable.

\begin{figure}[tbh]
\centerline{\includegraphics[width=7.8cm,height=10.5cm,clip]{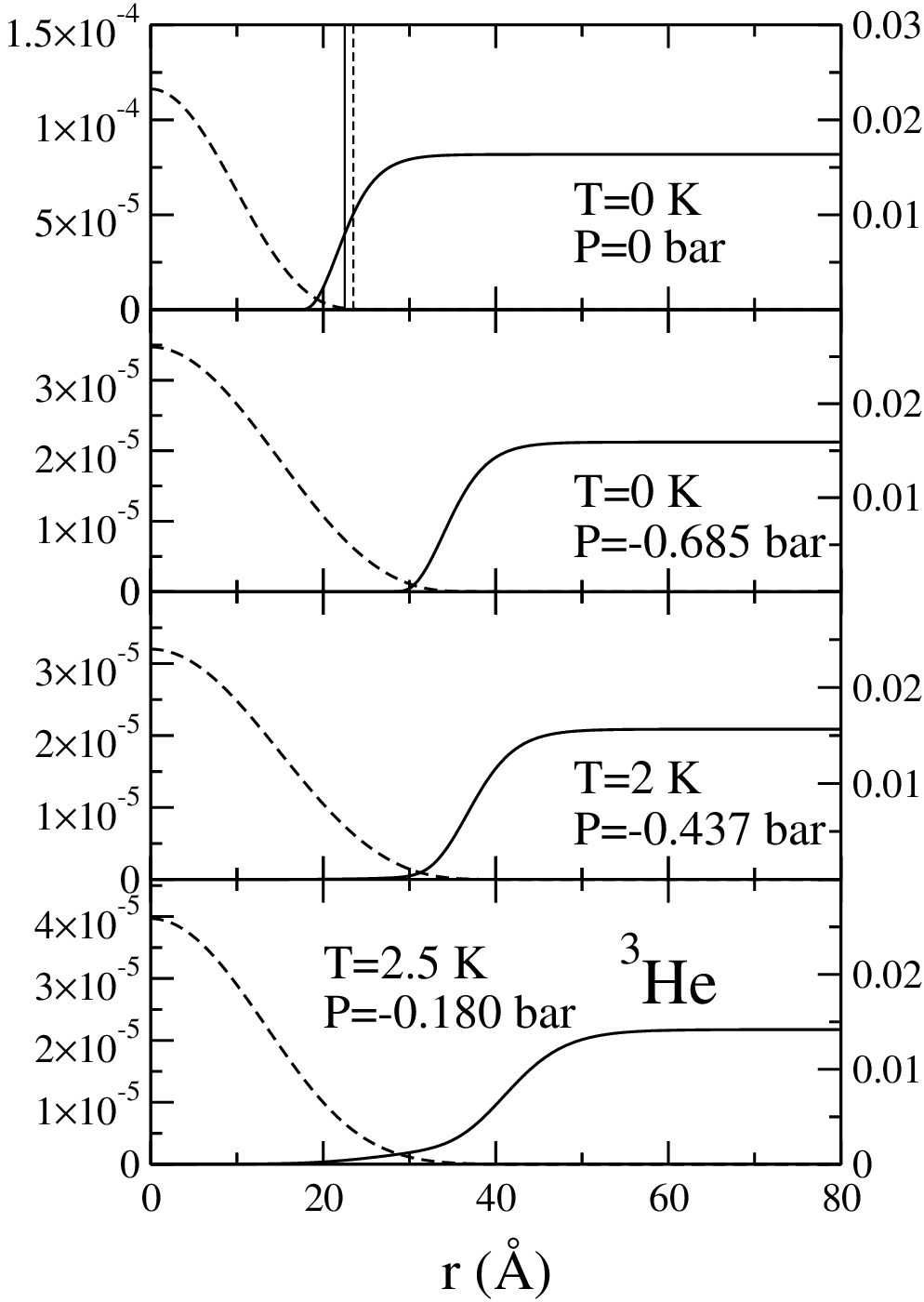} }
\caption[]{
$^3$He density profiles in \AA$^{-3}$ (solid lines, right scale)
and excess electron squared wave functions $|\Psi|^2$ in
\AA$^{-3}$  (dashed lines, left scale),
as a function of radial distance $r$ (\AA).
The top panel shows the stable bubble at $T=0$ K,
$P=0$ bar. The other three panels show
the near-to-unstable electron bubble for several $(P,T)$
values. In the upper panel, the vertical thin dashed line indicates
the radius of the capillary model bubble ($R_{min} =  23.5$ \AA) ,
and the vertical thin solid
line indicates the radius at which the helium density equals $\rho_b/2$
[$R(\rho_b/2) = 22.5$ \AA, with $\rho_b = 0.0163$ \AA$^{-3}$].
 }
\label{fig8}
\end{figure}

We represent in Fig. \ref{fig8} several $^3$He density profiles
and excess electron squared wave functions $|\Psi|^2$.
The top panel shows the stable bubble at $T=0$ K,
$P=0$ bar. The other three panels
display the near-to-unstable electron bubble for several $(P,T)$
values. Comparing with Fig. \ref{fig3}, it can be seen that $^3$He
electron bubbles are more diffuse than $^4$He electron bubbles.

\begin{figure}[tbh]
\centerline{\includegraphics[width=10cm,height=6.5cm,clip]{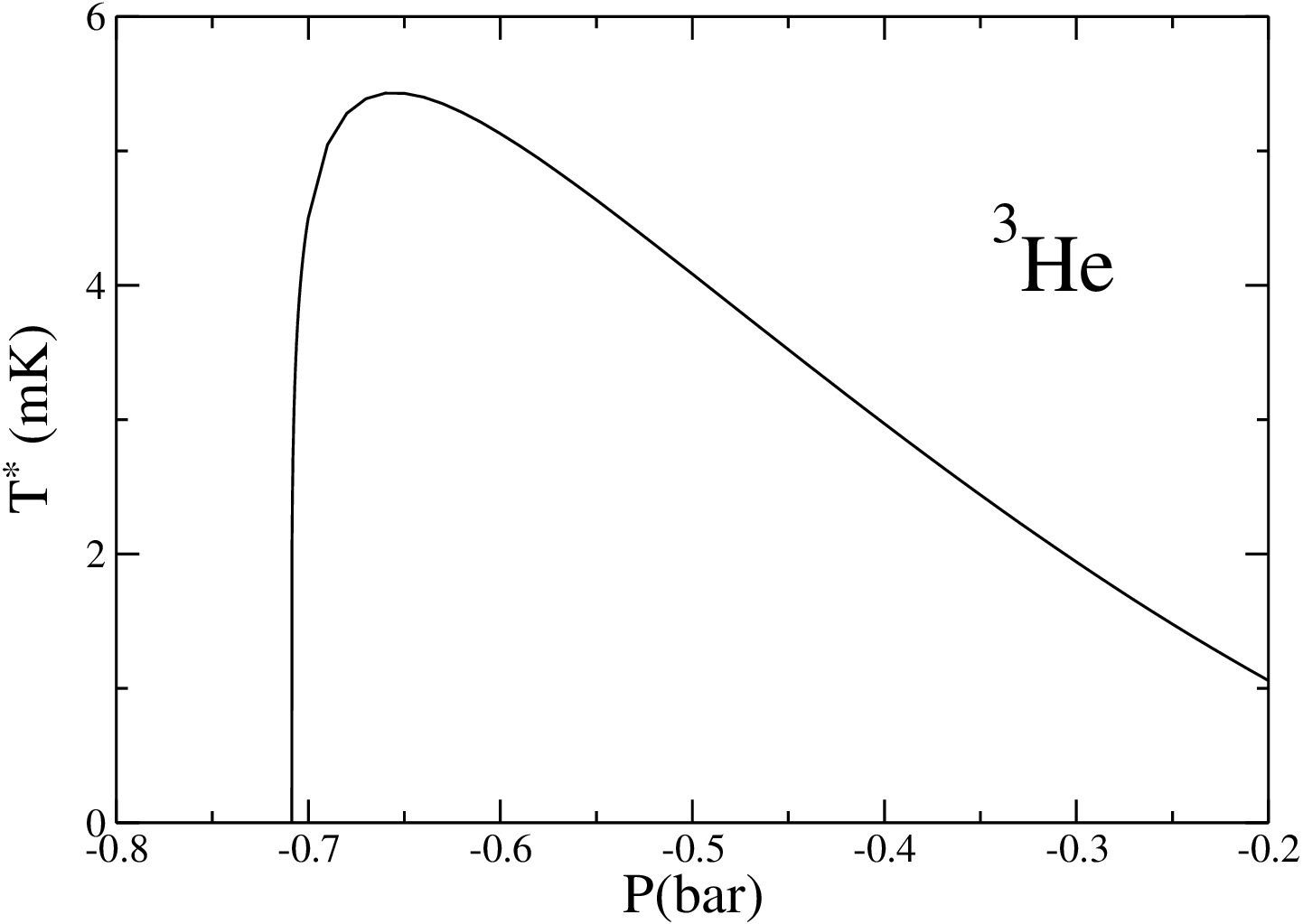} }
\caption[]{
Crossover temperature $T^*$ (mK) as a function of $P$ (bar)
of an electron bubble in liquid $^3$He in the capillary model.
 }
\label{fig9}
\end{figure}

We have also obtained $T^*$ for $^3$He in the capillary model.
In spite of theoretical predictions that point to a crossover temperature
of the order of 100 mK in pure liquid $^3$He,\cite{Mar95,Gui97} recent
experiments have not found such a crossing\cite{Cau02}, indicating that
superfluid coherence might play a role in quantum cavitation. Yet,
using the expressions given in Sec. 2 we have obtained $T^*(P)$ for $^3$He
and show it in Fig. \ref{fig9}. It can be seen that the maximum of
$T^*(P)$ is roughly half the value we have obtained for $^4$He in the
capillary model. Proceeding as in the case of $^4$He and assuming the
same experimental parameteres, we have found that the value of $T^*$
that would correspond to this situation is about 2.3 mK. This estimate is
close to the normal-to-superfluid transition temperature in $^3$He,
below which our DF method does not apply as it assumes that $^3$He is in
the normal phase.

\section{SUMMARY}

We have thoroughly addressed electron bubble explosions
in liquid helium. Our approach is
based on the application of  finite temperature density functionals
succesfully used to describe cavitation in liquid $^4$He and $^3$He.
This approach is fully
selfconsistent and unbiased by numerical artifacts, and to our
knowledge, it is the only one applied in a  wide range of
temperatures.

We have compared our results with experiments and have found that
our calculations are either in agreement with the experimental
data,\cite{Cla98,Su98} or compatible with them
if they have error bars similar to these
attributed to other cavitation processes in liquid helium.

We have used a realistic electron-helium interaction and have
tested another approach based on the use of a
simpler interaction. We have found that in
spite of the fairly large electron bubbles involved in the process,
at high temperatures
the capillary model fails to yield quantitative results, overestimating
the critical pressures. We attribute this to the
`rigidity' of the bubble configurations which is inherent to the capillary
approach. Whereas these
are serious drawbacks for nanoscopic bubbles, they are expected 
not to have a sizeable influence for microscopic multielectron 
bubbles.\cite{Tem03}

We have also used the density functional results in conjunction with a
functional integral method to obtain the thermal to quantum crossover
temperature. This approach has led in the past to a correct description
of the same process in pure liquid $^4$He. In the present
case, the crossover temperature turns out to be very small, about 6 mK.

Finally, we want to stress the suitability of the DF
approach to quantitatively address electron bubbles in
liquid He. This might encourage one to investigate other problems
like the infrared spectrum of the electron bubble in liquid helium, 
and the effect of quantized vortices pinned to excess electrons
on the critical cavitation pressure.
It has been argued\cite{Cla98a} that the
rising of $P_u$ below $T=1$ K could be attributed to the
presence of quantized vortices.
However, only simple models have been used to study their effect, and
the agreement with experiment is only qualitative. In the case of the 
infrared spectrum of electron bubbles, the DF approach might shed light on the
long-standing problem of how to understand the experimental results of Grimes and
Adams\cite{Gri90} on the $1s-1p$ and $1s-2p$ electron transition energies without
using unjustifiable pressure dependences of the helium surface tension
within the capillary model.

\section*{ACKNOWLEDGMENTS}
We would like to thank F. Ancilotto, F. Caupin,  M. Guilleumas,
D. M. Jezek, H. Maris and F. Toigo  for useful discussions.
This work has been performed under grants BFM2002-01868 from DGI (Spain) and
2000SGR00024 from  Generalitat de Catalunya.

\end{document}